\documentclass[prl,aps,twocolumn,preprintnumbers,amsmath,amssymb,superscriptaddress,showpacs]{revtex4}

\usepackage{graphicx}
\usepackage{dcolumn}
\usepackage{bm}
\usepackage{epstopdf}

\newcount\driver
\newcount\bozza

scaled\magstep1                  
scaled\magstep1 
 
scaled\magstep1                 \font\indbf=cmbx10 scaled\magstep2

scaled \magstep2

{\count255=\time\divide\count255 by 60
\xdef\hourmin{\number\count255}
        \multiply\count255 by-60\advance\count255 by\time
   \xdef\hourmin{\hourmin:\ifnum\count255<10 0\fi\the\count255}}

\let\a=\alpha \let\b=\beta    \let\g=\gamma     \let\d=\delta     \let\e=\varepsilon
  \let\h=\eta     \let\th=\vartheta      
\let\m=\mu    \let\n=\nu                      \let\r=\rho
\let\s=\sigma             
   \let\o=\omega     
 \let\D=\Delta       \let\L=\Lambda    \let\X=\Xi

\def\EE{{\cal E}}\def\VV{{\cal V}}
\def\WW{{\cal W}}
\def\BB{{\cal B}}
\def\RR{{\cal R}}\def\LL{{\cal L}}

\def\pp{{\bf p}}\def\xx{{\bf x}}
   
\def\yy{{\bf y}}\def\kk{{\bf k}}\def\nn{{\bf n}}

       \def\oo{{\underline \omega}}
\def\ee{{\underline \varepsilon}}

        \def\EE{\hbox{\msytw E}}

\let\==\equiv

\let\io=\infty
\let\0=\noindent

\def\*{{\hfill\break\null\hfill\break}}

\def\media#1{{\langle#1\rangle}}

\def\tilde#1{{\widetilde #1}}

\def\tende#1{\,\vtop{\ialign{##\crcr\rightarrowfill\crcr
             \noalign{\kern-1pt\nointerlineskip}
             \hskip3.pt${\scriptstyle #1}$\hskip3.pt\crcr}}\,}
\def\otto{\,{\kern-1.truept\leftarrow\kern-5.truept\to\kern-1.truept}\,}

\def\Tr{\rm Tr}

\def\wh#1{\widehat{#1}}
\def\hat#1{\wh{#1}}
\def\sqt[#1]#2{\root #1\of {#2}}

\def\bp{{\bar \ps}}

\def\EE{{\cal E}}\def\VV{{\cal V}}
\def\WW{{\cal W}}
\def\BB{{\cal B}}
\def\RR{{\cal R}}\def\LL{{\cal L}}

\def\T#1{{#1_{\kern-3pt\lower7pt\hbox{$\widetilde{}$}}\kern3pt}}
\def\VVV#1{{\underline #1}_{\kern-3pt
\lower7pt\hbox{$\widetilde{}$}}\kern3pt\,}
\def\W#1{#1_{\kern-3pt\lower7.5pt\hbox{$\widetilde{}$}}\kern2pt\,}

\def\indica{\leaders \hbox to 0.5cm{\hss.\hss}\hfill}
\def\guida{\leaders\hbox to 1em{\hss.\hss}\hfill}
\mathchardef\oo= "0521

\def\pp{{\bf p}}\def\xx{{\bf x}}
\def\yy{{\bf y}}\def\kk{{\bf k}}\def\nn{{\bf n}}

\def\oo{{\underline \omega}}

\def\qed{\raise1pt\hbox{\vrule height5pt width5pt depth0pt}}
 
  \def\bp{{\bar p}} 

\def\indic{\hbox{\raise-2pt \hbox{\indbf 1}}}

\def\ins#1#2#3{\vbox to0pt{\kern-#2 \hbox{\kern#1 #3}\vss}\nointerlineskip}

\newdimen\xshift \newdimen\xwidth \newdimen\yshift
\newcount\griglia

\def\insertplot#1#2#3#4#5#6{%
\xwidth=#1pt \xshift=\hsize \advance\xshift by-\xwidth \divide\xshift by 2%
\begin{figure}[ht]
\vspace{#2pt} \hspace{\xshift}
\begin{minipage}{#1pt}
#3 \ifnum\driver=1 \griglia=#6
\ifnum\griglia=1 \openout13=griglia.ps \write13{gsave .2
setlinewidth} \write13{0 10 #1 {dup 0 moveto #2 lineto } for}
\write13{0 10 #2 {dup 0 exch moveto #1 exch lineto } for}
\write13{stroke} \write13{.5 setlinewidth} \write13{0 50 #1 {dup 0
moveto #2 lineto } for} \write13{0 50 #2 {dup 0 exch moveto #1
exch lineto } for} \write13{stroke grestore} \closeout13
\includegraphics{griglia.ps} \fi
\includegraphics{#4.ps}\fi%
\ifnum\driver=2 \fi
\end{minipage}
\caption{#5}
\end{figure}
}

\newdimen\shift \shift=-1.5truecm
\def\lb#1{%
\ifnum\bozza=1
\label{#1}\rlap{\hbox{\hskip\shift$\scriptstyle#1$}}
\else\label{#1} \fi}

\def\be{\begin{equation}}
\def\ee{\end{equation}}
\def\bea{\begin{eqnarray}}\def\eea{\end{eqnarray}}
\def\bean{\begin{eqnarray*}}\def\eean{\end{eqnarray*}}
\def\bfr{\begin{flushright}}\def\efr{\end{flushright}}
\def\bc{\begin{center}}\def\ec{\end{center}}
\def\bal{\begin{align}}\def\eal{\end{align}}
\def\ba#1{\begin{array}{#1}} \def\ea{\end{array}}
\def\bd{\begin{description}}\def\ed{\end{description}}
\def\bv{\begin{verbatim}}\def\ev{\end{verbatim}}
\def\nn{\nonumber}
\def\Halmos{\hfill\vrule height10pt width4pt depth2pt \par\hbox to \hsize{}}
\def\pref#1{(\ref{#1})}

\def\ins#1#2#3{\vbox to0pt{\kern-#2 \hbox{\kern#1 #3}\vss}\nointerlineskip}

\newdimen\xshift \newdimen\xwidth \newdimen\yshift
\newcount\griglia

\def\insertplot#1#2#3#4#5#6{%
\xwidth=#1pt \xshift=\hsize \advance\xshift by-\xwidth \divide\xshift by 2%
\begin{figure}[ht]
\vspace{#2pt} \hspace{\xshift}
\begin{minipage}{#1pt}
#3 \ifnum\driver=1 \griglia=#6
\ifnum\griglia=1 \openout13=griglia.ps \write13{gsave .2
setlinewidth} \write13{0 10 #1 {dup 0 moveto #2 lineto } for}
\write13{0 10 #2 {dup 0 exch moveto #1 exch lineto } for}
\write13{stroke} \write13{.5 setlinewidth} \write13{0 50 #1 {dup 0
moveto #2 lineto } for} \write13{0 50 #2 {dup 0 exch moveto #1
exch lineto } for} \write13{stroke grestore} \closeout13
\includegraphics{griglia.ps} \fi
\includegraphics{#4.ps}\fi%
\ifnum\driver=2 \fi
\end{minipage}
\caption{#5}
\end{figure}
}

\newdimen\shift \shift=-1.5truecm
\def\lb#1{%
\label{#1}\rlap{\hbox{\hskip\shift$\scriptstyle#1$}}
\else\label{#1} \fi}

\def\be{\begin{equation}}
\def\ee{\end{equation}}
\def\bea{\begin{eqnarray}}\def\eea{\end{eqnarray}}
\def\bean{\begin{eqnarray*}}\def\eean{\end{eqnarray*}}
\def\bfr{\begin{flushright}}\def\efr{\end{flushright}}
\def\bc{\begin{center}}\def\ec{\end{center}}
\def\bal{\begin{align}}\def\eal{\end{align}}
\def\ba#1{\begin{array}{#1}} \def\ea{\end{array}}
\def\bd{\begin{description}}\def\ed{\end{description}}
\def\bv{\begin{verbatim}}\def\ev{\end{verbatim}}
\def\nn{\nonumber}
\def\Halmos{\hfill\vrule height10pt width4pt depth2pt \par\hbox to \hsize{}}
\def\pref#1{(\ref{#1})}

\usepackage{amsmath}
\usepackage{amsfonts}
\usepackage{amssymb}
\usepackage{slashbox}
\usepackage{epstopdf}

scaled\magstep1

\let\a=\alpha \let\b=\beta  \let\g=\gamma  \let\d=\delta
\let\e=\varepsilon
  \let\h=\eta   \let\th=\theta  
\let\m=\mu    \let\n=\nu             \let\r=\rho
\let\s=\sigma     
   \let\o=\omega
 \let\D=\Delta  \let\L=\Lambda \let\X=\Xi

\def\EE{{\cal E}} \def\VV{{\cal V}}
 \def\WW{{\cal W}}
 
\def\RR{{\cal R}}\def\LL{{\cal L}}

 \def\pp{{\bf p}}
 \def\xx{{\bf x}} \def\yy{{\bf y}} 
\def\kk{{\bf k}}

\def\nn{\nonumber}

\def\\{\hfill\break}
\def\={:=}
\let\io=\infty
\let\0=\noindent
\def\media#1{{\langle#1\rangle}}

\def\tende#1{\,\vtop{\ialign{##\crcr\rightarrowfill\crcr\noalign{\kern-1pt
    \nointerlineskip} \hskip3.pt${\scriptstyle #1}$\hskip3.pt\crcr}}\,}
\def\otto{\,{\kern-1.truept\leftarrow\kern-5.truept\to\kern-1.truept}\,}

\def\wh{\widehat}
\def\to{\rightarrow}

\def\qed{\hfill\raise1pt\hbox{\vrule height5pt width5pt depth0pt}}

\def\be{\begin{equation}}
\def\ee{\end{equation}}
\def\bp{\begin{pmatrix}}
\def\ep{\end{pmatrix}}
\def\bea{\begin{eqnarray}}
\def\eea{\end{eqnarray}}
\def\nn{\nonumber}
\def\pref#1{(\ref{#1})}

\def\lb{\label}

\def\Tr{\mathrm{Tr}}

scaled\magstep1                  
scaled\magstep1 
 
scaled\magstep1                 \font\indbf=cmbx10 scaled\magstep2

scaled \magstep2

{\count255=\time\divide\count255 by 60
\xdef\hourmin{\number\count255}
        \multiply\count255 by-60\advance\count255 by\time
   \xdef\hourmin{\hourmin:\ifnum\count255<10 0\fi\the\count255}}

\let\a=\alpha \let\b=\beta    \let\g=\gamma     \let\d=\delta     \let\e=\varepsilon
  \let\h=\eta     \let\th=\vartheta      
\let\m=\mu    \let\n=\nu                      \let\r=\rho
\let\s=\sigma             
   \let\o=\omega     
 \let\D=\Delta       \let\L=\Lambda    \let\X=\Xi

\def\EE{{\cal E}}\def\VV{{\cal V}}
\def\WW{{\cal W}}
\def\BB{{\cal B}}
\def\RR{{\cal R}}\def\LL{{\cal L}}

\def\pp{{\bf p}}\def\xx{{\bf x}}
   
\def\yy{{\bf y}}\def\kk{{\bf k}}\def\nn{{\bf n}}

       \def\oo{{\underline \omega}}
\def\ee{{\underline \varepsilon}}

        \def\EE{\hbox{\msytw E}}

\let\==\equiv

\let\io=\infty
\let\0=\noindent

\def\*{{\hfill\break\null\hfill\break}}

\def\media#1{{\langle#1\rangle}}

\def\tilde#1{{\widetilde #1}}

\def\tende#1{\,\vtop{\ialign{##\crcr\rightarrowfill\crcr
             \noalign{\kern-1pt\nointerlineskip}
             \hskip3.pt${\scriptstyle #1}$\hskip3.pt\crcr}}\,}
\def\otto{\,{\kern-1.truept\leftarrow\kern-5.truept\to\kern-1.truept}\,}

\def\Tr{\rm Tr}

\def\wh#1{\widehat{#1}}
\def\hat#1{\wh{#1}}
\def\sqt[#1]#2{\root #1\of {#2}}

\def\bp{{\bar \ps}}

\def\EE{{\cal E}}\def\VV{{\cal V}}
\def\WW{{\cal W}}
\def\BB{{\cal B}}
\def\RR{{\cal R}}\def\LL{{\cal L}}

\def\T#1{{#1_{\kern-3pt\lower7pt\hbox{$\widetilde{}$}}\kern3pt}}
\def\VVV#1{{\underline #1}_{\kern-3pt
\lower7pt\hbox{$\widetilde{}$}}\kern3pt\,}
\def\W#1{#1_{\kern-3pt\lower7.5pt\hbox{$\widetilde{}$}}\kern2pt\,}

\def\indica{\leaders \hbox to 0.5cm{\hss.\hss}\hfill}
\def\guida{\leaders\hbox to 1em{\hss.\hss}\hfill}
\mathchardef\oo= "0521

\def\pp{{\bf p}}\def\xx{{\bf x}}
\def\yy{{\bf y}}\def\kk{{\bf k}}\def\nn{{\bf n}}

\def\oo{{\underline \omega}}

\def\qed{\raise1pt\hbox{\vrule height5pt width5pt depth0pt}}
 
  \def\bp{{\bar p}} 

\def\indic{\hbox{\raise-2pt \hbox{\indbf 1}}}

\def\ins#1#2#3{\vbox to0pt{\kern-#2 \hbox{\kern#1 #3}\vss}\nointerlineskip}

\newdimen\xshift \newdimen\xwidth \newdimen\yshift
\newcount\griglia

\def\insertplot#1#2#3#4#5#6{%
\xwidth=#1pt \xshift=\hsize \advance\xshift by-\xwidth \divide\xshift by 2%
\begin{figure}[ht]
\vspace{#2pt} \hspace{\xshift}
\begin{minipage}{#1pt}
#3 \ifnum\driver=1 \griglia=#6
\ifnum\griglia=1 \openout13=griglia.ps \write13{gsave .2
setlinewidth} \write13{0 10 #1 {dup 0 moveto #2 lineto } for}
\write13{0 10 #2 {dup 0 exch moveto #1 exch lineto } for}
\write13{stroke} \write13{.5 setlinewidth} \write13{0 50 #1 {dup 0
moveto #2 lineto } for} \write13{0 50 #2 {dup 0 exch moveto #1
exch lineto } for} \write13{stroke grestore} \closeout13
\includegraphics{griglia.ps} \fi
\includegraphics{#4.ps}\fi%
\ifnum\driver=2 \fi
\end{minipage}
\caption{#5}
\end{figure}
}

\newdimen\shift \shift=-1.5truecm
\def\lb#1{%
\ifnum\bozza=1
\label{#1}\rlap{\hbox{\hskip\shift$\scriptstyle#1$}}
\else\label{#1} \fi}

\def\be{\begin{equation}}
\def\ee{\end{equation}}
\def\bea{\begin{eqnarray}}\def\eea{\end{eqnarray}}
\def\bean{\begin{eqnarray*}}\def\eean{\end{eqnarray*}}
\def\bfr{\begin{flushright}}\def\efr{\end{flushright}}
\def\bc{\begin{center}}\def\ec{\end{center}}
\def\bal{\begin{align}}\def\eal{\end{align}}
\def\ba#1{\begin{array}{#1}} \def\ea{\end{array}}
\def\bd{\begin{description}}\def\ed{\end{description}}
\def\bv{\begin{verbatim}}\def\ev{\end{verbatim}}
\def\nn{\nonumber}
\def\Halmos{\hfill\vrule height10pt width4pt depth2pt \par\hbox to \hsize{}}
\def\pref#1{(\ref{#1})}

\def\ins#1#2#3{\vbox to0pt{\kern-#2 \hbox{\kern#1 #3}\vss}\nointerlineskip}

\newdimen\xshift \newdimen\xwidth \newdimen\yshift
\newcount\griglia

\def\insertplot#1#2#3#4#5#6{%
\xwidth=#1pt \xshift=\hsize \advance\xshift by-\xwidth \divide\xshift by 2%
\begin{figure}[ht]
\vspace{#2pt} \hspace{\xshift}
\begin{minipage}{#1pt}
#3 \ifnum\driver=1 \griglia=#6
\ifnum\griglia=1 \openout13=griglia.ps \write13{gsave .2
setlinewidth} \write13{0 10 #1 {dup 0 moveto #2 lineto } for}
\write13{0 10 #2 {dup 0 exch moveto #1 exch lineto } for}
\write13{stroke} \write13{.5 setlinewidth} \write13{0 50 #1 {dup 0
moveto #2 lineto } for} \write13{0 50 #2 {dup 0 exch moveto #1
exch lineto } for} \write13{stroke grestore} \closeout13
\includegraphics{griglia.ps} \fi
\includegraphics{#4.ps}\fi%
\ifnum\driver=2 \fi
\end{minipage}
\caption{#5}
\end{figure}
}

\newdimen\shift \shift=-1.5truecm
\def\lb#1{%
\label{#1}\rlap{\hbox{\hskip\shift$\scriptstyle#1$}}
\else\label{#1} \fi}

\def\be{\begin{equation}}
\def\ee{\end{equation}}
\def\bea{\begin{eqnarray}}\def\eea{\end{eqnarray}}
\def\bean{\begin{eqnarray*}}\def\eean{\end{eqnarray*}}
\def\bfr{\begin{flushright}}\def\efr{\end{flushright}}
\def\bc{\begin{center}}\def\ec{\end{center}}
\def\bal{\begin{align}}\def\eal{\end{align}}
\def\ba#1{\begin{array}{#1}} \def\ea{\end{array}}
\def\bd{\begin{description}}\def\ed{\end{description}}
\def\bv{\begin{verbatim}}\def\ev{\end{verbatim}}
\def\nn{\nonumber}
\def\Halmos{\hfill\vrule height10pt width4pt depth2pt \par\hbox to \hsize{}}
\def\pref#1{(\ref{#1})}

\begin{document}

\title{Interacting Weyl semimetals on a lattice}

\author{Vieri Mastropietro}
\affiliation{
Universit\`a degli Studi di Milano, Via Cesare Saldini 50,  
Milano, Italy}
\begin{abstract}
Electron-electron interactions in a Weyl semimetal are rigorously investigated in a lattice model by non perturbative methods. 
The absence of quantum phase transitions is proved
for interactions not too large and short ranged.  The anisotropic Dirac cones persist with angles (Fermi velocities) 
renormalized by the interaction, and with generically shifted Fermi points.
As in graphene, the optical conductivity shows universality properties:
it is equal to the massless Dirac fermions one with
renormalized velocities, up to
corrections which are subdominant in modulus.
\end{abstract}
\maketitle

\section{I. Introduction}

There is a wide interest in materials with an emerging description in terms of Dirac particles,
like graphene \cite{G}, a {\it bidimensional} system which owes much of its remarkable features to the fact that its Fermi surface is
point-like and its charge carries approximatively behave as $D=2+1$ Dirac particles.
Among most remarkable properties of graphene is the universality of the optical conductivity \cite{N}, which appears to be independent from microscopic details. This is in contrast with the Fermi velocity which is instead renormalized by the interaction \cite{E}. 

In recent times several proposal for {\it three dimensional} materials, called {\it Weyl semimetals}, with point-like Fermi surfaces 
and an emerging description
in terms of $D=3+1$ Dirac fermions has bee advanced, see \cite{V,W,B,Y}.  First experiments reporting the realization  of such materials have been reported \cite{L,L1,L3}. In real systems
 the charge carriers surely interact and it is important to understand 
how the interaction modifies the physical properties. In particular, is the interaction able to open a gap in the spectrum or to produce a quantum phase transition?
Does the optical conductivity show
universality properties as in the case of graphene?

We will consider the three dimensional lattice model 
introduced in \cite{DC} (see also \cite{VV}), with nearest and next to nearest neighbor hopping and with a magnetic flux density, such that the flux over each surface of the unit cell is vanishing.
As in \cite{Ha}, the effect of the magnetic flux is to decorate the hopping terms with certain phases, and for certain choices it can be shown,
see \cite{DC}, that for suitable densities the Fermi surface is given by two points and that the dispersion relation is approximately conical, so that an effective description in terms of Dirac fermions in $D=3+1$ dimensions is possible. To this model we add a short-ranged interaction between fermions and we analyze how the 
physical properties are modified by the interaction. 

An important point of our analysis is that we will fully take into account
the effects of {\it non linear bands} and of the lattice and we will provide {\it non perturbative}  results via a
convergent multiscale expansion. 
There are good reason why it is worthwhile to do that. 
In the case of graphene (and Weyl semimetals are their 3D analogue),
the use of an 
effective continuum Dirac model has led to ambiguities, due to the fact that  
the Dirac description has spurious ultraviolet divergences (absent in the lattice model)
which require a suitable regularization procedure, {\it e.g.} dimensional or momentum regularization.
In particular, the graphene conductivity in the case of long range interactions appears strongly dependent on the approximations used
(that is, on how the truncation of the series expansion is done) and on the choice of the regularization; this has generated a debate \cite{a1}-\cite{a8} which is still not
completely  settled, despite there is certain agreement that the sensitivity to regularizations imposes the physical realization of the cut-off provided by the lattice \cite{a6a},\cite{a6},\cite{a7}. 
The situation is much clearer in the case of graphene with short range interactions. 
Perturbative computations using different regularizations led to different conclusions, but at the end
a rigorous result \cite{GM}, \cite{GMPcond} has settled the question, showing that in the Hubbard model on a honeycomb lattice
in the half filled band case 
the optical conductivity is {\it universal}
in the zero frequency limit; all the possible interaction correction exactly cancel out.
This was one of the very few cases in which an universality result has been deduced from first principles starting from a microscopic many body hamiltonian; it is {\it non perturbative} (a lowest order computation could never establish exact universality) and requires
to take into account the irrelevant terms (they give finite contributions which are essential for cancellations).

In this paper we will analyze by similar non perturbative methods a lattice Weyl semimetal with short range interactions. 
We can rigorously exclude, at least inside the radius of convergence of the renormalized expansions, 
the presence of mass generation or quantum instabilities; the (anisotropic) Dirac cones persist also in presence of the interaction, whose only effect is to change the values of the angles. 
As consequence of lattice Ward Identities, the optical conductivity displays universality properties, as it 
is equal to the free one with
renormalized velocities, up to
corrections which are subdominant in modulus.
In particular the optical conductivity has a real part vanishing as the frequency $O(\o)$ \cite{H} and an imaginary part vanishing as $O(\o\log|\o|)$ \cite{RL}; 
we show, at a non perturbative level, that in the interacting case the conductivity is equal to the free one with renormalized Fermi velocities, up to $O(\o)$ corrections.

The paper is organized as follows. In \S II the model is presented and the main results are stated. In 
\S III the lattice Ward Identities are derived. In \S IV we set up our exact Renormalization Group analysis
and in \S V such analysis is combined with Ward Identities to get our main results; finally in \S VI we state our conclusions.

\section{II. Lattice Weyl semimetals}

We consider a sublattice $\L_A=\L$ with side $L$ given by the points $\vec x=(\vec n \vec \d)$, 
with $\vec\d_1=(1,0,0)$, $\vec\d_2=(0,1,0)$, $\vec\d_3=(0,0,1)$; we consider also 
a sublattice $\L_B$ whose points are $\vec x+\vec\d_+$ with $\vec\d_{+}={{\vec \d_1}+{\vec d_2}\over 2}$ and $\vec\d_{-}={{\vec \d_1}-{\vec d_2}\over 2}$.
Introducing
fermionic creation and annihilation operators 
$(a^\pm_{\vec x}, b^\pm_{\vec x+{\vec \d_+}})$, 
the  Hamiltonian is $H=H_1+H_2+H_3+U V$ where
$H_1$ contains the nearest-neighbor planar $AB$  hopping
\bea
&&H_1={1\over 2}\sum_{\vec x\in \L}\{[
 -i t( a^+_{\vec x} b^-_{\vec x+\vec\d_+}+b^+_{\vec x+\vec\d_+}a_{\vec x+2\vec\d_+})+H.c.]+\nn\\
&&[ t( a^+_{\vec x} b^-_{\vec x-\vec\d_-}-b^+_{\vec x-\vec\d_-}a_{\vec x-2\vec\d_-})+
H.c.]\}
\eea
while $H_1$ contains  $AA$ or $BB$ hopping
\bea
&&H_2={1\over 2}\sum_{\vec x\in \L}\{[
t_{\perp }(a^+_{\vec x} a^-_{\vec x+\vec \d_3}-b^+_{\vec x+\vec\d_+}b_{\vec x+\vec\d_++\vec\d_3})+
H.c.]\nn\\
&&
-t'\sum_{i=1,2}[(a^+_{\vec x} a^-_{\vec x+{\vec \d_i}}-
b^+_{\vec x+\vec\d_+} b^-_{\vec x+\vec\d_++{\vec\d_i}})
+H.c.]\}\nn
\eea
and $H_3$ takes into account the on site energy difference between the two sublattices
\be
H_3={\m\over 2}\sum_{\vec x\in \L}(a^+_{\vec x} a^-_{\vec x}-b^+_{\vec x+\vec\d_+} b^-_{\vec x+\vec\d_+})
\ee
Finally the interaction is given by
\be
V=\sum_{\vec x,\vec y} v(\vec x-\vec y) 
[a^+_{\vec x} a^-_{\vec x}+
b^+_{\vec x+\vec \d_+} b^-_{\vec x+\vec \d_+}][
 a^+_{\vec y} a^-_{\vec y}+ b^+_{\vec y+\vec \d_+} b^-_{\vec y+\vec \d_+}
]\ee
where $v(\vec x)$ is a quasi-local interaction.
It is also convenient to define
$a^+_{\vec x}={1\over L^3}\sum_{\vec k} e^{i\vec k\vec x}\hat a^+_{\vec k}$ and 
$b^+_{\vec x+\vec\d_+}={1\over L^3}\sum_{\vec k} e^{i\vec k(\vec x+\vec\d_+)}\hat b^+_{\vec k}$, where 
$\vec k={2\pi\over L}\vec n$.

The {\it current} is defined as usual via the {\it Peierls
substitution}, by modifying  the hopping parameter along the bond
$(\vec x,\vec x+ \vec\d)$ as  $ U_{\vec x,\vec x+\vec \d}(\vec A )=e^{i e\int_0^{1} \vec \d\cdot \vec A(\vec x+ s \vec \d) ds}$, where $e$ is the electric charge and
$\vec A=(A_1,A_2,A_3)$ is the vector electromagnetic field; the modified Hamiltonian is 
\be H(\vec A)=H_1(\vec A)+H_2(\vec A)+H_3+ U V
\ee
where
\begin{widetext}
\bea
&&H_1(A)={1\over 2}\sum_{\vec x\in \L}\{[
 -i t( a^+_{\vec x} U_{\vec x,\vec x+\vec\d_+}b^-_{\vec x+\vec\d_+}+b^+_{\vec x+\vec\d_+} U_{\vec x+\vec\d_+,\vec x+2\vec\d_+} b^-_{\vec x+2\vec\d_+})+
it ( b^+_{\vec x+\vec \d_+} U_{\vec x+\vec\d_+,\vec x}a^-_{\vec x}+
a^+_{\vec x+2\vec\d_+} U_{\vec x+2\vec\d_+,\vec x+\vec\d_+} b^-_{\vec x+\vec\d_+})
]\label{d1}\\
&&+[ t( a^+_{\vec x}U_{\vec x,\vec x-\vec\d_-} b^-_{\vec x-\vec\d_-}-b^+_{\vec x-\vec\d_-}U_{\vec x-\vec\d_-,\vec x-2\vec\d_-} a_{\vec x-2\vec\d_-})+
( b^+_{\vec x-\vec\d_-}U_{\vec x-\vec\d_-,\vec x} a^-_{\vec x}-a^+_{\vec x-2\vec\d_-}U_{\vec x-2\vec\d_-,\vec x-\vec\d_-} b_{\vec x-\vec\d_-})]\nn\\
&&H_2(A)={1\over 2}\sum_{\vec x\in \L}\{[
t_{\perp }(a^+_{\vec x} U_{\vec x,\vec x+\vec\d_3}
a^-_{\vec x+\vec \d_3}-b^+_{\vec x+\vec\d_+}  U_{\vec x+\vec\d_+,\vec x+\vec\d_++\vec\d_3}b_{\vec x+\vec\d_++\vec\d_3})+
a^+_{\vec x+\vec\d_3} U_{\vec x+\vec\d_3,\vec x}
a^-_{\vec x}-\nn\\
&&
b^+_{\vec x+\vec\d_++\vec\d_3}  U_{\vec x+\vec\d_++\vec\d_3,\vec x+\vec\d_+}b_{\vec x+\vec\d_+})]\nn\\&&
-t'\sum_{i=1,2}[(a^+_{\vec xi}   U_{\vec x,\vec x+\vec\d_i}
a^-_{\vec x+\vec\d_i}-
b^+_{\vec x+\vec\d_++\vec\d_i} 
U_{\vec x+\vec\d_+,\vec\d_j,\vec x+\vec\d_+}
b^-_{\vec x+\vec\d_++\vec\d_i})
+
a^+_{\vec x}   U_{\vec x,\vec x+\vec\d_i}
a^-_{\vec x+\vec\d_i}-
b^+_{\vec x+\vec\d_+} 
U_{\vec x+\vec\d_+,\vec x+\vec\d_++\vec\d_i}
b^-_{\vec x+\vec\d_++\vec\d_i})]\}\nn
\eea
\end{widetext}
The paramagnetic lattice current is given by \be
j_\pm (\vec p) =-{\partial H(\vec A)\over \partial A_{\pm,\vec p}}|_0,\quad
j_3 (\vec p) =-{\partial H(\vec A)\over \partial A_{3,\vec p}}|_0
\ee if
$A_{\pm}={A_1\pm A_2\over 2}$. Using 
the notation $\int d\kk={1\over\b L^3}\sum_{\kk}$ one gets,  if $\hat\psi^\pm_{\vec k}=(a^\pm_{\vec k},b^\pm_{\vec k})$
\bea
&&\hat j_{+;\vec p}=e\int d\vec k \hat\psi^+_{\vec k+\vec p}[w_{a,+}(\vec k,\vec p)\s_1+w_{b,+}(\vec k,\vec p) \s_3]
\hat\psi^-_{\vec k}\label{cu1}\nn\\
&&\hat j_{-;\vec p}=e\int d\vec k \hat\psi^+_{\vec k+\vec p}[w_{a,-}(\vec k,\vec p)\s_2+w_{b,-}(\vec k,\vec p) \s_3]
\hat\psi^-_{\vec k}\nn\\
&&\hat j_{3,\vec p}=e\int d\vec k \psi^+_{\vec k} w_{3}(\vec k,\vec p) \s_3
\hat\psi^-_{\vec k}
\eea
where
\bea
&&w_{a,\pm }(\vec k,\vec p)={i \over 2}t \h_\pm(\vec p) (e^{i(\vec k+\vec p)\d_\pm}+e^{-i\vec k \d_\pm})\nn\\
&&w_{b,\pm}(\vec k,\vec p)=i{t'\over 2} \sum_{i=1,2} \h_i(\vec p) [e^{i(\vec k+\vec p)\vec\d_i}-e^{-i\vec k\vec\d_i}]\nn\\
&&w_{3}(\vec k,\vec p)= -i {t_\perp\over 2} \h_3(\vec p)(e^{i(\vec k+\vec p)\vec\d_3}-e^{- i \vec k\vec\d_3})
\eea
with $\h_i(\vec p)={1-e^{-i \vec p\vec\d}\over i\vec p\vec \d}=1+O(\vec p)$. Similarly the diamagnetic current is defined as
$j^Â_{\vec p}={\partial^2 H(\vec A)\over \partial \hat A_{\vec p} \partial \hat A_{\vec p}} |_0$.
The density operator is defined as $\r_{\vec x}=a^+_{\vec x} a^-_{\vec x}+b^+_{\vec x+\vec\d_+} b^-_{\vec x+\vec\d_+}$.
Moreover $\s_0=I$ and $\s_i$, $i=1,2,3$ are Pauli matrices.
If $O_{\xx}=e^{x_{0}H}O_{\vec
x_i}e^{-x_{0}H}$, with $\xx=(x_{0},\vec x)$, we denote by
\be
\media{O^{(1)}_{\xx_1}\cdots O^{(n)}_{\xx_n}}_{\b}=\lim_{L\to\io}\X^{-1}\Tr\{e^{-\b H}{\bf T}(O^{(1)}_{\xx_1}\cdots O^{(n)}_{\xx_n})\}\ee
where $\X=\Tr\{e^{-\b H}\}$ and ${\bf T}$ is the operator of
fermionic time ordering; moreover we denote by
$\media{O^{(1)}_{\xx_1};\cdots ;O^{(n)}_{\xx_n}}_{\b}$ the
corresponding {\it truncated expectations} and by
$\media{O^{(1)}_{\xx_1};\cdots ;O^{(n)}_{\xx_n}}$ their zero temperature limit.

We will particularly interested in the {\it two-point Schwinger function} $\media{\psi^-_\xx\psi^+_\yy}$
and in the 
conductivity, which is defined via Kubo formula.
Denoting by $\media{\hat j_{i;\pp}
\hat j_{i;-\pp}}$, $\pp=(\o,\vec p)$, the Fourier transform of 
$\media{\hat j_{i;\xx}
\hat j_{i;\yy}}$, the conductivity is  
\be \s_{ii}(i\o)= -\lim_{\b\to\io}\lim_{\b\to\io}{1\over\o}\Big[\media{\hat j_{i;\o,0};
\hat j_{i;-\o,0}}_\b+\D_{i}\Big]\label{cond}\ee
where $\D_{i}$ is the diamagnetic contribution. We will consider the {\it optical} conductivity, obtained taking the limit ${\b\to\io}$ at $\o$ fixed and then $\o\to 0$, corresponding to the regime
$a>>\o>>\b^{-1}$, where $a=1$ is the lattice mesh. 

In the {\it non interacting case} $U=0$ the properties of the system can be easily computed.
The dispersion relation is given by $|\EE(\vec k)|$ where
\bea
&&\EE(\vec k)=t\sin (\vec k\vec\d_+)\s_1+t\sin (\vec k\vec \d_-)\s_2+\nn\\
&&\s_3(\m+t_\perp \cos k_3- {1\over 2}t'(\cos k_1+\cos k_2))
\eea
We assume $t,t_\perp,t'$ positive, $\m+t'>t_{\perp}$ and $0<\m-t'<t_{\perp}$ so that the dispersion relation $|\EE(\vec k)|$ vanishes only for $\vec k=(0,0,\pm p_F)$, where $t_\perp \cos p_F=\m-t'$.
The $2$-point Schwinger function $\media{\psi^-_\xx\psi^+_\yy}|_{U=0}\equiv g(\xx-\yy)$ is given by
\begin{widetext}
\be
g(\xx)=\int d\kk e^{i\kk\xx} 
\begin{pmatrix} &-i k_0+t_\perp(\cos k_3-\cos k_F)+E(\vec k)& t(\sin k_+-i\sin k_-)\\
&
t(\sin k_++i \sin k_-)& 
-i k_0-t_{\perp}(\cos k_3-\cos k_F)-E(\vec k)
\end{pmatrix}^{-1}\label{prop1}
\ee
\end{widetext}
with $E(\vec k)=t'(\cos k_+\cos k_--1)$. 
Denoting by $\pp_F^\pm=(0,0,0,\pm p_F)$ we get,
for $\kk$ close to $\pp_F^\pm$, $k_3'=k_3\pm p_F$
\bea
\hat g(\kk)\sim 
\begin{pmatrix}&-i k_0\pm v_{3,0} k'_{3}
& v_{\pm, 0}(k_+-i k_-)\\
&
v_{\pm 0}(k_++i k_-)& 
-i k_0\mp v_{3,0} k'_3
\end{pmatrix}^{-1}\nn
\eea
with
\be
v_{\pm ,0}=t\quad\quad  v_{3,0}=t_\perp \sin p_F
\ee
The 2-point function is asymptotically close to the one of massless free Dirac fermions in $D=3+1$ dimensions, up to the fact that 
the light velocity is replaced by two Fermi velocities, respectively $v_{\pm,0}$ for the directions $1,2$ and $v_{3, 0}$ for the 3-direction. The conductivity computed through \pref{cond} is given by, $l=+,-,3$ 
\be
\s_{ll}(i\o)|_{U=0}=e^2 {v_{l,0}^2\over (v_{\pm,0})^2 v^0_{3, 0}}\s_{ii,weyl}(i\o)+O(\o)
\ee
where $\s_{ii,weyl}(i\o)$ is the conductivity of Weyl fermions with light velocity $c=1$
%
%
which is vanishing in modulus as $\o\to 0$ as
\be
\s_{ll,weyl}(i\o)=O(\o\log |\o|)\ee
By analytic continuation $i\o\to \o+\i\e$ one can verify that the real part of the non interacting conductivity vanishes as $O(\o)$
\cite{H} while the imaginary part vanishes as $O(\o\log |\o|)$ \cite{RL}.

Our main result can be summarized by the following Theorem.
\vskip.3cm
{\bf Theorem} {\it If $\m+t'>t_{\perp}$ and $0<\m-t'<t_{\perp}$
and $U$ small enough
the Schwinger functions are analytic in $U$ uniformly as $L,\b\to\io$. 
$\media{\hat\psi^-_\kk\hat\psi^+_\kk}$ is singular at 
$\kk=(0,0,0,0,\pm p_F)$ , $p_F=\cos^{-1}(\m-t')t^{-1}_{\perp}+O(U)$
and close to such points
\be \media{\psi^-_\kk\psi^+_\kk}\sim
{1\over Z}
\begin{pmatrix}&-i k_0\pm v_{3} k'_{3}
& v_{\pm}(k_+-i k_-)\\
&
v_{\pm}(k_++i k_-)& 
-i k_0\mp v_{3} k'_3
\end{pmatrix}^{-1}\label{pal}
\ee
with
\be
v_{\pm}= v_{\pm,0}+a_\pm U+O(U^2), \quad v_{3}= v_{3,0}+a_3 U+O(U^2)\label{xx1}
\ee
$a_\pm, a_{3}$ given by \pref{fon11} below
and $Z=1+O(U^2)$. Finally the conductivity is given by
\be
\s_{ll}(i\o)=e^2 {v_{l}^2\over (v_{\pm})^2 v_{3}}\s_{ii,weyl}(i\o)+R(\o))\label{xx}
\ee
with $|R(\o)|\le C|\o|$.}
\vskip.3cm
The proof is based on a multiscale expansion, whose convergence follows from results in constructive QFT about infrared QED$_{3+1}$ \cite{M}.
This is therefore one of the very rare systems in three dimensions in which many body effects can be understood with full mathematical rigor.
In particular, the above result excludes non-perturbative phenomena like quantum instabilities or gap generation; the asymptotic behavior of the two point function is the same as in the non interacting case, 
and the role of the interaction is to modify the value of the Fermi velocities and of the wave function renormalization.  The optical conductivity  
is equal to the one of Weyl fermions with
renormalized velocities, up to
subdominant corrections. Either the renormalization of the Fermi velocities \pref{xx1}  and the identity 
\pref{xx} could be experimentally verified; for instance
the conductivity and the Fermi velocities can be 
measured by independent experiments (and in different realizations of lattice semimetals), and the above result says that 
the combination $\s_{ll}(i\o){(v_{\pm})^2 v_{3}\over v_l^2}$
is universal as $\o \to 0$.

\section{III. Ward Identities}

%
%

The physical observables can be expressed as usual in terms of Grassmann integrals. We denote by 
$\psi^\pm_{\xx}=(a^\pm_\xx,b_{\xx+{\bf d}_+})$, $\xx=(x_0,x)$, ${\bf d}=(0,\vec \d)$ a set of Grassmann variables; with abuse of notation, we denote them by the same symbol as the fermionic fields.
As we expect that the location of the Fermi points will be in general modified by 
the presence of the interaction, we find convenient
to fix it to its non interacting value by replacing $\m$ with 
$t_\perp cos p_F+\n$, where $\n$ is a counterterm to be suitably chosen as function of $U$; a non vanishing $\n$ means that the location of the Fermi points is shifted.
We introduce the {\it generating functional}, ${\bf A}=(A_0,\vec A)$
\be
e^{\WW({\bf A},\phi^+,\phi^-)}=\int P(d\psi)e^{\VV(\psi)+\BB(A)+(\psi,\phi)}
\ee
where $P(d\psi)$ is the fermionic integration with propagator \pref{prop1} and
 $\VV$ is the interaction
\be
\VV=\n N+U V
\ee
where, if $\int d\xx= \int dx_0\sum_{\vec x}$
\bea
&&N=\int d\xx\psi^+_{\xx}\s_0
\psi^-_{\xx}\\
&&V=\int d\xx d\yy 
 v(\xx-\yy)(\psi^+_{\xx}\s_0\psi^-_{\xx})(\psi^+_{\yy}\s_0\psi^-_{\yy})\nn
\eea
if $v(\xx)=\d(x_0) v(x)$. Moreover 
$(\psi,\phi)=\int d\xx [\psi^+_{\xx}\s_0\phi^-_{\xx}+\psi^-_{\xx}\s_0\phi^+_{\xx}]$
and  
\be
\BB({\bf A},\psi)=\int d\xx A_0(\xx) \psi^+_{\xx}\s_0\partial_0 \psi^-_\xx+\BB_1(A,\psi)+\BB_2(A,\psi)
\ee
where the explicit expression of $\BB_1(A,\psi)$ and $B_2(A,\psi)$ is obtained by $H_1(A)$ and $H_2(A)$ \pref{d1} by replacing $\sum_{\vec x}$
with $\int dx_0 \sum_{\vec x}$, the Fermi operators $a_{\vec x},b_{\vec x}$ with the Grassmann variables $a_{\xx},b_{\xx}$ and 
$U_{\xx,\xx+{\bf d}}(A)=e^{i e\int_0^{1} \vec \d \cdot \vec A(\xx+ s {\bf d}) ds}$. 
As in lattice gauge theory, by the change of variables $a^\pm_{\xx} \to e^{\mp i e \a_{\xx}}a_\xx^\pm$,
$b^\pm_{\xx+{\bf d_+}} \to e^{\mp i e \a_{\xx+{\bf d}_+} } b^\pm_{\xx+{\bf d}_+}$ 
and using the
relation $U_{\xx,\xx+{\bf d}}(A)=e^{i e \a_{\xx+{\bf d}}-i e \a_\xx}$ one obtains
\be
W({\bf A}+{\bf \partial} \a,\phi^+ e^{i e \a},\phi^- e^{-i e \a})=W({\bf A},\phi^+,\phi^-)\label{cc}
\ee
where the fact that the Jacobian of the transformation is equal to $1$ has be exploited; due  
to the presence of the lattice, no anomalies are present. From \pref{cc} we get the following identity
\be
W({\bf A}+{\bf \partial}\a,\phi^+ e^{i e \a},\phi^- e^{-i e \a})=0
\ee
from which by differentiating with respect to the external fields ${\bf A},\phi$ an infinite number of {\it Ward Identities}
connecting correlation functions is obtained.

In particular, if $\pp=(\o,\vec p)$
\be
-i \o <\hat\r_\pp; \hat\psi_{\kk}\hat\psi_{\kk+\pp}>+\sum_{j=\pm,3}p_j 
<\hat j_{j,\pp};
\hat\psi_{\kk}\hat\psi_{\kk+\pp}>=0\label{wi1}
\ee
where the currents $j_+,j_-,j_3$ are given by \pref{cu1}.
%
%
Similarly we can derive equation for the current-current correlation
\bea
&&-i \o \media{\hat\r_\pp;\hat\r_{-\pp}}+\sum_{l=\pm, 3}
p_l \media{\hat j_{l,\pp}; \hat \r_{-\pp}}=0\label{w2}\\
&&-i \o \media{\hat\r_\pp ;\hat j_{i,-\pp}}+
\sum_{l=\pm,3}p_l \media{\hat j_{l,\pp}; \hat j_{i,-\pp}}
+p_i \D_i=0\nn
\eea
where we have used that 
\bea
&&
{\partial^2 \WW(A)\over \partial A_{i,\pp} \partial A_{l,-\pp}}|_0=
\media{\hat j_{i,\pp}; \hat j_{l,\-\pp}}\quad i\not =j\nn\\
&&{\partial^2 \WW(A)\over \partial A_{i,\pp} \partial A_{i,-\pp}}|_0=
\media{\hat j_{i,\pp}; \hat j_{i,-\pp}}+\D_i
\eea
From \pref{w2} we get
the following equality
\be
{\o^2\over p_i^2} \media{\hat\r_{\bar\pp};\hat\r_{-\bar\pp}}=
\media{\hat j_{i;\bar\pp} ;\hat j_{i;-\bar\pp}}+\D_i\label{w4}
\ee
where $\bar\pp$ is obtained from $\pp$ setting $p_{l}=0$, $l\not=(0,i)$.
From the above equation we get, differentiating with respect to $\o$
\be
{2 \o\over p_i^2} \media{\hat\r_{\bar\pp};\hat\r_{-\bar\pp}}+
{\o^2\over p_i^2} \partial_\o\media{\hat\r_{\bar\pp};\hat\r_{\bar\pp}}
=\partial_\o \media{\hat j_{i;\bar\pp}; \hat j_{i;-\bar\pp}}\label{w5}
\ee
from which we get
\be
\media{\hat j_{i;\bar\pp} ;\hat j_{i;-\bar\pp}}|_{\o=0}+\D_i=0\quad\quad
\partial_\o \media{\hat j_{i;\bar\pp}; \hat j_{i;-\bar\pp}}|_{\o=0}=0\label{w10}
\ee
From \pref{w10} we see that the properties of the conductivity \pref{cond}
depend crucially on the continuity and differentiability 
of the Fourier transform of the current-current correlations. 
Indeed if $\media{\hat j_{i,\pp}; \hat j_{i,-\pp}}$ is continuous in $\pp$ then
from the first of \pref{w10} we get 
\be \s_{ll}(i\o)= -\frac1{\o} \Big[\media{\hat j_{l;\o,0}
;\hat j_{l;-\o,0}}-\media{\hat j_{i;0,0};
\hat j_{l;0,0}}\Big]\label{31}
\ee
and if the derivative is continuous then is vanishing by the second of \pref{w10}.
In the non interacting case $\media{j_{i;\xx};
\hat j_{i;\xx}}$ decays as $O(|\xx-\yy|^{-6})$ at large distance; therefore the Fourier transform is continuous and with continuous first derivative, hence 
$\s^{0}_{ii}(
\o)$ vanishes as $\o\to 0$. However the second derivative is not continuous, and this explains the logarithmic correction in the free conductivity.
In the following sections we will determine the continuity properties of the current correlation in presence of interaction in order to deduce the conductivity properties of the interacting system.

\section{IV. Renormalization Group analysis}

We describe now how the generating functional can be analyzed using 
a multiscale Renormalization Group methods.
It is important to stress that the method is {\it exact}; in particular, contrary to usual field theoretical Renormalization Group analysis,
the {\it irrelevant terms} (including lattice effects and non linear bands) are fully taken into account; as we said this is crucial in view of the sensitivity 
to cut-offs of certain quantities, like the conductivity, in Dirac metals.
The starting point is the following decomposition of the propagator \pref{prop1}
\be
\hat g(\kk)=\hat g^{(u.v.)}(\kk)+\hat g^{(i.r.)}(\kk)
\ee
where $\hat g^{(u.v.)}(\kk)=(1-\chi(\kk))\hat g(\kk)$, $\hat g^{(i.r.)}(\kk)=\chi(\kk)\hat g(\kk)$, where $\chi(\kk)$ is a smooth
cut-off function vanishing for $\sqrt{(k_0^2+|\EE(\vec k)|^2)}< a_0$,
with $a_0={t-t'\over 4 t_\perp}$, and $\hat g^{(u.v.)}(\kk)=(1-\chi(\kk))\hat g(\kk)$.
Note that the support of $\chi(\kk)$ consists in two disconnected regions around the two Fermi points $\pp_F^\pm=(0, 0,0,\pm p_F)$; therefore we can write 
\be\hat g^{(i.r.)}(\kk)=\sum_{\e=\pm }\hat g^{(\le 0)}_\e(\kk)\ee with $\hat g^{(\le 0)}_\e(\kk)$ non vanishing only in a region around $\pp^\e_F$; in coordinate space 
\be
g(\xx)=g^{(u.v.)}(\xx)+\sum_{\e=\pm} e^{i\e\pp_F\xx} g^{(\le 0)}_{\xx,\e}(\xx)
\ee
with $g^{(\le 0)}_{\xx}(\xx)=\int d\kk' e^{i\kk'\xx} \hat g^{(\le 0)}_\e(\kk'+\e\pp_F)$.

We use the {\it addition property} of Grassman integrals writing the generating functional as,
calling $V({\bf A},\psi,\phi)=\VV(\psi)+\BB({\bf A},\psi)+(\psi,\phi)$ 
\be
e^{\WW}=\int P(d\psi) e^{V}=\int P(d\psi^{(u.v.)})\prod_{\e=\pm}P(d\psi_\e^{(\le 0)})e^{V({\bf A},\psi,\phi)}
\ee
where $P(d\psi^{(u.v.)})$ and  $P(d\psi_\e^{(i.r.)})$
are the fermionic integration with propagator $g^{(u.v.)}(\xx)$ and $g^{(\le 0)}_\e(\xx)$
respectively and $\psi^\pm_{\xx}=\psi^{(u.v.)}_\xx+\sum_{\e=\pm} e^{\pm i\e \pp_F\xx} \psi^{(\le 0)\pm}_{\xx,\e}$.
The first step of the (exact)
Renormalization Group analysis is based on the integration of the ultraviolet component
\be
e^{\WW}=
\int \prod_{\e=\pm}P(d\psi_\e^{(\le 0)}) e^{V^{(0)}({\bf A},\psi^{(\le 0)},\phi)}
\ee
where (in the $\phi=0$ for definiteness)
\be
V^{(0)}({\bf A},\psi^{(\le 0)},0)
=\sum_{n,m}W^{(0)}_{n,m,l}[\prod_{l=0}^n \psi^{\s_l(\le 0)}_{\e_l,\xx_l}][\prod_{k=0}^m A_{k\yy_k}]
\ee
The integration of the field $\psi^{\le 0}$ is performed in a multiscale fashion, writing 
$\psi^{\pm(\le 0)}_{\e,\xx}=\sum_{h=-\io}^0 \psi^{\pm(h)}_{\e,\xx}$. where  $\psi^{\pm(h)}_{\e}$
lives on a shell of momenta scales closer and closer $O(2^h)$) from $\pp^\e_F$. After the integration
of $\psi^{(0)},..,\psi^{(h+1)}$ one gets
\be
e^{\WW}=\int \prod_\e P(d\psi_\e^{(\le h)}) e^{V^{(h)}({\bf A},\sqrt{Z_h}\psi^{(\le h)},\phi)}\label{h1}
\ee
where $P(d\psi_\e^{(\le h)})$ has propagator given by
\begin{widetext}
\be
g^{(\le h)}(\xx)=\int d\kk' e^{i\kk'\xx}{\chi_h(\kk)\over Z_h} 
\begin{pmatrix}&-i k_0+v_{3,h}(\cos (k'_3+\e {\vec p_F})-\cos k_F)+E(\vec k)& 
v_{\pm ,h}(\sin k_+-i\sin k_-)\\
&
v_{\pm,h}(\sin k_++i \sin k_-)& 
-i k_0-v_{3,h}(\cos (k'_3+\vec p_F)-\cos k_F)-E(\vec k)
\end{pmatrix}^{-1}\label{cond2}
\ee
\end{widetext}
with $\chi_h(\kk')$ is $1$ for $\sqrt{(k_0^2+|\EE(\vec k'+\e\vec p_F)|^2)}< a_0 2^h$
and  $0$ for $\sqrt{(k_0^2+|\EE(\vec k+\e\vec p_F)|^2)}>a_0 2^{h+1}$. 
Finally $V^{(h)}$
is similar to $V^{(0)}$, with kernels $W^{(h)}_{n,m}$ and fields $\psi^{\s(\le h)}_{\e,\xx}$. 
$Z_h$ and $v_{\pm, h}, v_{3,h}$ are respectively the wave function renormalization and the Fermi velocities at the momentum scale $|\kk-\e\pp_F|\sim 2^h$.
The scaling dimension of the kernels $W^{(h)}_{n,m}$ is given by
\be
D=4-{3\over 2}n-m
\ee
According with the usual RG terminology, the terms with negative scaling dimension are called {\it irrelevant},
the terms with vanishing dimension are called {\it marginal} and the terms with positive dimension are the {\it relevant}
terms. As in the case of graphene (in which $D=3-n-m$ )
, the terms with four or more fermionic fields are irrelevant; this is sharp contrast to what happens for Dirac fermions in $1+1$ dimension 
($D=2-n/2-m$) in which the quartic terms are marginal.

We define a {\it localization operator} acting on the kernels $\hat W_{n,m}(\kk')$ with non negative dimension in the following way;
$\LL=0$ except for $(2,1)$ and $(4,0)$; in that case, $\m=0,+,-,3$ 
\begin{widetext}
\bea
&&\LL\int d\kk d\pp \hat W^{(h)}_{\m;2,1}(\kk'+\e\pp_F,\pp)\hat A_{\m,\pp} \hat\psi^+_{\kk'+\pp+\e \pp_F,\e}\hat\psi^-_{\kk+\e\pp_F,\e}=\int d\kk d\pp \hat W^{(h)}_{\m;2,1}(\e\pp_F,0)A_{\m,\pp}\hat\psi^+_{\kk+\pp+\e\pp_F,\e}\hat\psi^-_{\kk+\e\pp_F,\e}\quad \a=\pm 3\nn\\
&&\LL\int d\kk  \hat W_{2,0}^{(h)}(\kk'+\e\pp_F)\hat\psi^+_{\kk'+\e\pp_F,\e}\hat\psi^-_{\kk'+\e\pp_F}=\int d\kk 
[\hat W_{2,0}^{(h)}(\e\pp_F)+\kk'{\bf\partial}\hat W^{(h)}_{2,0}(\e\pp_F)]\hat\psi^+_{\kk+\e\pp_F,\e}\hat\psi^-_{\kk'+\e\pp_F,\e}
\eea
\end{widetext}
Note that there are no bilinear terms $\psi^+_{\e}\psi_{-\e}$; in the case $m=0$ this follows from
conservation of momentum and when $m=1$ this follows from the fact that we assume $\pp$ small.
Note that the propagator verify the following symmetry properties, calling $\kk^*=(k_0,-k_1,-k_2,k_3)$
\bea
&&\hat g_{1,1}(\kk)=\hat g_{1,1}(\kk^*)\quad \hat g_{2,2}(\kk)=\hat g_{2,2}(\kk^*)
\nn\\
&&\hat g_{1,2}(\kk)=-\hat g_{1,2}(\kk^*)\quad \hat g_{2,1}(\kk)=-\hat g_{2,1}(\kk^*)\label{p1}
\eea
Morever the kernels of the currents verify
\bea
&&w_{a,\pm}(\vec k,0)=w_{a,\pm}(\vec k^*,0)\quad w_{b,\pm}(\vec k,0)=-w_{b,\pm}(\vec k^*,0)\nn\\
&&w_{3}(\vec k,0)=w_{3}(\vec k^*,0)
\eea
By using the above symmetry properties it is easy to check that
\begin{enumerate}
\item The non diagonal terms $\hat W^{(h)}_{2,0}(\e\pp_F)$ are vanishing by \pref{p1} as they contain an odd number of non diagnal propagators, by \pref{p1}.
\item The non diagonal terms contributing to $\partial_0 \hat W^{(h)}_{2,0}(\e\pp_F)$
or $\partial_3 \hat W^{(h)}_{2,0}(\e\pp_F)$
are vanishing as they contain an odd number of non diagonal contributions; similarly
diagonal terms contributing to $\partial_1\hat  W^{(h)}_{2,0}(\e\pp_F)$
or $\partial_2 \hat W^{(h)}_{2,0}(\e\pp_F)$.
\item The diagonal contributions to $\hat W_{\pm;2,1}(\e\pp_F,0)$ are vanishing; indeed the terms
containing $w_{b,\pm}$ contains an even number of non diagonal propagators, hence they are vanishing;
the terms containing $w_{a,\pm}$ contains instead an odd number of non diagonal propagators.
\item The non diagonal contributions to 
$\hat W^{(h)}_{3;2,1}(\e\pp_F,0)$ are vanishing as they contain an odd number of non diagonal propagators.
\end{enumerate}
We write
\be
\LL V^{(h)}=\LL\VV^{(h)}+\LL\BB^{(h)}
\ee
%
%
%
%
and we shift all the terms in $\LL\VV^h$ in the free integration 
except the renormalization of the Fermi points; we rescale the fields 
$\psi^{(\le h)}\to {\sqrt{Z_{h-1}}\over\sqrt{Z_h}}\psi^{(\le h)}$
so that finally \pref{h1}
becomes
\bea
&&\int \prod_\e \tilde P(d\psi_\e^{(\le h)}) e^{\n_{h-1} Z_{h-1} F^{(h)}_\n(\sqrt{Z_{h-1}}\psi^{(\le h)})}\times\nn\\
&&e^{\LL\BB^h({\bf A},\sqrt{Z_{h-1}}\psi^{(\le h)}),\phi  ) +\RR V^{(h)}}
\eea
where $F_\n^{(h)}=\sum_{\e=\pm}\int d\kk \hat\psi^+_{\kk'+\e\pp_F,\e}\s_0\hat\psi^-_{\kk'+\e\pp_F,\e}$
and $\tilde P(d\psi_\e^{(\le h)})$ has a propagator similar to $g^{(\le h)}$ \pref{cond2} with $Z_h,v_{\pm, h},v_{3, h}$ replaced by
$Z_{h-1},v_{\pm,h-1},v_{3, h-1}$. 
Moreover
\bea
&&\LL B^{(h)}({\bf A},\sqrt{Z_{h-1}}\psi^{(\le h)},0)= \sum_{\e=\pm}
\int d\kk' d\pp \hat\psi^{(\le h)}_{\kk'+\e\pp_F,\e}\nn\\
&&[Z_{0,h}A_0(\pp)\s_0
+Z_{+,h}A_+(\pp)\s_1\nn\\
&&
+Z_{-,h}A_-(\pp)\s_2+
+\e Z_{3,h}
\s_3]\hat\psi^{(\le h)}_{\kk'+\pp+\e\pp_F,\e}
\eea
%
Note the natural emergence in the relevant part of the source of {\it relativistic}
Dirac currents, quite simpler with respect to the lattice currents \pref{cu1}; the irrelevant terms are stored in $\RR V^{(h)}$ and will contribute to the renormalization of the physical quantities (even more, the renormalization in only due to the irrelevant terms). The renormalized parameters obey by construction to iterative flow beta function equations 
\bea
&&\n_{h-1}={Z_h\over Z_{h-1}}(\g \n_h+ \g^{-h} \hat W_{2}^{(h)}(\e\pp_F))\nn\\
&&{Z_{h-1}\over Z_h}=1+\partial_0 \hat W_{2}^h(\e\pp_F)\label{fl}\\
&&v_{\a, h-1}={Z_h\over Z_{h-1}}(v_{\a,h}+\partial_\a \hat W^{(h)}_{2,0}(\e\pp_F))\quad\a=\pm,3\nn\\
&&{Z^{(\m)}_{h-1}\over Z_{h-1}}={Z^{(\m)}_h\over Z_{h-1}}[1+\hat W_{2,1}^{(h)}(\e\pp_F)]\quad \m=0,\pm,3
\eea
Finally we can write
\be
\tilde P(d\psi_\e^{(\le h)})=P(d\psi_\e^{(\le h-1)})P(d\psi_\e^{(h)})
\ee
and integrate the field $\psi^{(h)}$  so that the procedure can be iterated.

Assume that $Z_h, v_{\a,h}$ remain close to their value for $h=0$ and that $\n_h\sim O(U 2^{\th h})$ for some $0<\th<1$ (what will be proven inductively below). The single scale propagator corresponding to $P(d\psi_\e^{(h)})$ has a faster than any power decay; that is for any $N$
\be
|g^{(h)}(\xx)|\le 2^{3h}{C_N\over 1+[2^h |\xx|]^N}\label{fg}
\ee
By adapting a result for $QED_{4+1}$ in \cite{M} one can prove that
the kernels
$W^{(h)}_{n,m}$ in $V^{(h)}$
are
analytic in $U$ and decay super-polynomially in the relative
distances on scale $2^{-h}$; in particular, for all $0<\th<1$  they satisfy
the bounds
\be
{1\over L^3\b}\int d{\underline\xx}|W^{(h)}_{n,m}|\le C |U| 2^{(4-{3\over 2}n-m)h}2^{\th h}\label{aa2}
\ee
which are {\it non-perturbative}, i.e., they
are based on the {\it convergence} of the expansion for the
kernels $W^{(h)}$. Such estimate is obtained by exploiting the
anticommutativity properties of the Grassmann variables, via a
determinant expansion and the use of the {\it Gram-Hadamard
inequality} for determinants. Even if the number of Feynman graphs is $O(n!^2)$, 
no factorials appear in the r.h.s. of \pref{aa2}; this implies convergence and analyticity in $U$.
Note that in addition to the factor 
$2^{(4-{3\over 2}n-m)h}$ corresponding to the
scaling dimension there is a dimensional gain 
$2^{\th h}$  due to the irrelevance of the effective 
electron-electron interaction: every contribution in perturbation theory 
involving an effective scattering in the infrared is suppressed
thanks to the irrelevance of the kernels with four or more legs; the only terms with two legs are $O(2^{\th h})$ as well. 
This dimensional gain is analogous to the one found in 
super-renormalizable theories such as $\phi^4_2$ or $\phi^4_3$, thanks to the 
(exponentially fast) vanishing of the effective scattering term.

By using the bound \pref{aa2} in \pref{fl} we see that the beta function of the last three terms is $O(2^{\th h})$; therefore solving 
by iterations we get that the limiting values 
\bea
&&Z_{h}\to Z=1+O(U^2)\label{aa}\\
&&
v_{3,h}\to v_{3}=t_\perp\sin(p_F)+a_{3} U+O(U^2)
\nn\\
&&v_{\pm, h-1}\to v_{\pm}=t+a_\pm U+O(U^2)\nn\\
&&
Z_{0,h}\to Z_0=1+O(U^2)\nn\\
&&Z_{\pm ,h}\to Z_\pm=t+b_\pm U+O(U^2)\nn\\
&&Z_{3 ,h}\to Z_3=t_\perp\sin p_F+b_3 U+O(U^2)\nn
\eea
where
\bea
&&a_{3}=\int d\kk v(\kk) \partial_3 g_{11}(\kk)\label{fon11}\nn\\
&&a_{\pm}=\int d\kk v(\kk) \partial_\pm g_{12}(\kk)\\
&&b_{3}=\int d\kk v(\kk) w_3(\kk,0) [g_{1,1}(\kk) g_{1,1}(\kk)-
g_{2,1}(\kk) g_{1,2}(\kk)]\nn\\
&&b_\pm=\int d\kk v(\kk)[w_{a,\pm}(\vec k,0) (g_{11}(\kk)g_{2,2}(\kk)+g_{1,2}(\kk)g_{1,2}(\kk))
\nn\\
&&+w_{b,\pm}(\vec k,0)[g_{11}(\kk)g_{12}(\kk)-g_{1,2}(\kk)g_{22}(\kk))]\nn 
\eea
Regarding the flow of $\n_h$ we can write 
\be
\n_{h-1}=\g^{|h|}(\n+\sum_{k=0}^h \g^{k-2}b_\n^{(k)}]
\ee
where $b_\n^{(k)}=O(U\g^{\th k})$ by \pref{aa2}; therefore by choosing 
$\n=-\sum_{k=-\io}^{0} \g^{k-2}b_\n^{(k)}$ we get $\n_h=O(U \g^{\th h})$ and by an explicit computation
$\n=b U+O(U^2)$ where
\be
b=\int d\kk v(\kk)  g_{11}(\kk)
\ee
The single scale propagator can be written as 
\be
g^{(h)}_{\pm}(\xx)=g^{(h)}_{rel,\pm}(\xx)+r^{(h)}_\pm(\xx)
\ee
where
\be
\hat g^{(h)}_{rel,\pm}(\kk)={1\over Z_h}f_h(\kk)
\begin{pmatrix}&-i k_0\pm v_{3,h} k_{3}
& v_{\pm,h}(k_+-i k_-)\\
&
v_{\pm,h}(k_++i k_-)& 
-i k_0\mp v_{3,h} k'_3
\end{pmatrix}^{-1}\nn
\ee
where $f_h(\kk)$ selects momenta in a shell $O(2^h)$ 
and $r^{(h)}_\e$ decays faster in momentum spce, namely as 
\pref{fg} with $2^{3h}$ replaced by $2^{3h+\th h}$.

By construction the two point function is given by 
\be
\media{\psi^-_\xx\psi^+_{\bf 0}}=\sum_{h=-\io}^0\sum_{\e=\pm} e^{i\e\pp_F\xx}[g^{(h)}_\e(\xx)+A_h(\xx)]
\ee
where $|A_h(\xx)|\le C |U| 2^{3h} 2^{\th h}{C_N\over 1+[2^h |\xx|]^N}$.
By using \pref{aa2} and \pref{aa} we finally get \pref{pal}. In the same way, for $\m=0,\pm,3$ and $|\pp|<<|\kk-\e \pp_F|<<1$
\bea
&&<\hat j_{\m,\pp};\psi^+_{\kk}\hat\psi^-_{\kk+\pp}>=\\
&&e Z_{\m} \media{\hat\psi^-_\kk\hat\psi^+_{\kk}}\tilde\s_\m 
 \media{\hat\psi^-_{\kk+\pp}\hat\psi^+_{\kk+\pp}}\
(1+O(|\kk-\e\pp_F|^\th))\nn\label{aaaa}
\eea
where $Z_\m$ is given by \pref{aa} and $\tilde\s_+=\s_1$, $\tilde\s_-=\s_2$,
while $\tilde\s_0=\s_0$ and $\tilde\s_3=\s_3$.
Finally 
the current current correlation is given by, $i=\pm, 3$
\bea
&&
<j_{i,\xx}; j_{i,0}>=\\
&&e^2 [{Z_i\over Z}]^2
\sum_{\e=\pm }\int d\kk {\rm Tr}(\tilde \s_i g_{rel,\e}^{(\le 0)}(\kk)\tilde \s_i g_{rel,\e}^{(\le 0)}(\kk+\pp))
+H_i(\xx)\nn
\eea
where
\be
|H_i(\xx)|\le C |\xx|^{-6-\th}
\ee
and 
\be
\hat g^{(\le )}_{rel,\pm}(\kk)=\chi(\kk+\e\pp_F)
\begin{pmatrix}&-i k_0\pm v_{\perp} k_{3}
& v_\pm(k_+-i k_-)\\
&
v_\pm(k_++i k_-)& 
-i k_0\mp v_{\perp} k'_3
\end{pmatrix}^{-1}\nn
\ee
Therefore, the current current correlation can be written as sum of two terms; a dominant one, which decays as $O(|\xx|^{-6})$, and a rest with a faster decay, due to the improvement with respect to power counting in \pref{aa2}.

\section{V. Implications of Ward Identities and universality}

The velocities $v_{\pm}, v_{3}$, the wave function renormalization $Z$ and the renormalization of the currents $Z_\m$ are expressed by convergent series in $U$ and depend by all the microscopic details of the model. However the exact Ward Identity 
\pref{wi1} implies that they are not independent; by inserting \pref{pal} and \pref{aaaa}
in \pref{wi1} we get
\be
{Z_{\pm}\over Z }=v_\pm\quad\quad 
{Z_{3}\over Z}=v_3
\ee
The validity of the above identities can be verifies at lowest order; indeed
\be
a_{\perp}=b_{\perp}\quad\quad a_\pm=b_{\pm}
\ee
as they can be easily checked from the relations, $\pp_3=(0,\vec p\vec\d_3)$, 
$\pp_\pm=(0,\vec p\vec\d_\pm)$
\bea
&&g^{-1}(\kk)-g^{-1}(\kk+\pp_3)= p_3 w_{3}(\vec k,0) \s_3+O(\pp^2) \nn\\
&&g^{-1}(\kk)-g^{-1}(\kk+\pp_{\pm})=p_\pm [w_{a,\pm}(\vec k,0) \s_1\nn\\
&&+w_{b,\pm}(\vec k,0) \s_3]+O(\pp^2)
\eea
Therefore 
\bea
&&
<j_{i,\xx}; j_{i,0}>=\\
&&e^2 [v_l^2
\sum_{\e}\int d\kk {\rm Tr}(\tilde \s_i g_{rel,\e}^{(\le 0)}(\kk)\tilde \s_i g_{rel,\e}^{(\le 0)}(\kk+\pp))
+H_i(\xx)\nn
\eea
The first term is the same as the one 
in a non-interacting theory with velocities $v_\pm,v_3$ instead of $v_{0,\pm},v_{0,3}$.
In addition $H_i(\xx)$ has a faster decay, as for large distances behaves as 
$O(|\xx|^{-6-\th})$; therefore  the Fourier transform $\hat H_i(\pp)$ admits a continuous second derivative; hence
\bea
&&\hat H_i(\o,\vec 0)-\hat H_i(0,\vec 0)=\nn\\
&&\o \partial_\o \hat H_i(0,\vec 0)+{1\over 2}\o^2  \partial^2_\o \hat H_i(0,\vec 0)+\o^2 R_i(\o,\vec 0)\eea
with $R_i(0,\vec 0)=0$ and by  \pref{w10}  $\partial_\o H_i(0,\vec 0
)=0$; by using 
\pref{31} we get therefore \pref{xx}. In particular, the correction 
vanishes as $O(\o)$ with no extra log, and the dominant part
is equal to the free Dirac conductivity with renormalized velocities.





%


%
%

\section{VI.Conclusions}

We have rigorously shown that the interaction in a Weyl semimetal does not produce any quantum instability at weak coupling; its main effect is 
to modify the Fermi velocities and the wave number renormalization. 
Lattice Ward Identities and non perturbative bounds on the current-current correlations  imply that the optical conductivity  
is equal to the free one with
renormalized velocities, up to
subdominant corrections.
All our analysis is done assuming that the Fermi points are far enough; the convergence radius shrinks to zero when the two Fermi points coincide.
However extensions of the multiscale methods used in this paper are suitable to give results uniform in the distance between Fermi points. 
 In the Renormalization Group analysis  one has to consider two regimes, one in which the dispersion relation is essentially quadratic and another
at lower scales similar to the one treated here, and to use some gain from the first regime to compensate the vanishing of the Fermi velocity in the second regime expansion.
We plan to return on this in a future publication.



\end{document}